# Dislocations in (011)-oriented vertical Bridgman β-$Ga_2O_3$ substrates

Yongzhao Yao,[1,2,a)] Daiki Katsube,[2] Hirotaka Yamaguchi,[2] Yukari Ishikawa[2]

[1]Mie University, 1577 Kurimamachiya-cho, Tsu, Mie 514-8507, Japan
[2]Japan Fine Ceramics Center, 2-4-1 Mutsuno, Atsuta, Nagoya 456-8587, Japan

**Abstract:** Dislocation in (011)-oriented β-$Ga_2O_3$ substrates grown by the vertical Bridgman method was investigated using X-ray topography (XRT), combined with X-ray reticulography. Transmission XRT reveals dislocations lying on the (001) plane and extending along [010], forming arrays associated with domain boundaries. Dislocations on the (011) plane were also identified but differ from those responsible for line-shaped pits on (001) epilayers. Reflection XRT shows good agreement with transmission XRT and enables classification of dislocation types based on contrast features. Reticulography confirms domain boundaries with misorientation on the order of $10^{-5}$ rad, providing insight into defect formation relevant to epi-growth and device performance.

[a)]Author to whom correspondence should be addressed. Electronic mail: yao@icsdf.mie-u.ac.jp.
ORCID:0000-0002-7746-4204

$\beta$-$Ga_2O_3$ has emerged as a promising semiconductor for next-generation power devices owing to its outstanding physical properties, including an ultra-wide bandgap of ~4.5 eV and a high theoretical breakdown electric field of 8 MV/cm [1, 2]. These advantages make $\beta$-$Ga_2O_3$ particularly attractive for high-voltage and high-efficiency applications [3, 4], where thick, low-doped drift layers are required to realize superior device performance, especially high breakdown voltage [5, 6].

A critical issue in achieving high-quality $\beta$-$Ga_2O_3$ epitaxy is the choice of substrate orientation. For power device applications, epitaxial layers must satisfy several stringent requirements, including (i) a high growth rate to enable the formation of thick drift layers within a reasonable processing time, (ii) a smooth surface morphology without the need for post-growth treatments such as chemical mechanical polishing (CMP), (iii) a low background carrier concentration to achieve a high breakdown voltage, and (iv) a low density of extended defects such as dislocations and twins, some of which act as killer defects, leading to large leakage currents and premature breakdown under reverse bias. Achieving all of these requirements simultaneously remains challenging with conventional substrate orientations. To date, various orientations, including (100), (010), ($\bar{2}$01), (001), and (011), have been investigated; however, each exhibits intrinsic limitations. For instance, (100) and ($\bar{2}$01) substrates suffer from twin defects [7, 8], rendering them unsuitable for high-quality homoepitaxy. In contrast, although twin-related issues are less pronounced for (001) and (010) substrates, these orientations typically exhibit line-shaped pits and surface roughness after epitaxial growth [9], often necessitating CMP prior to device fabrication, thereby increasing processing costs.

Recently, the (011) orientation has attracted increasing attention as a promising alternative. It has been reported that the (011) plane can yield smooth, pit-free surfaces and a reduced density of critical defects, as dislocations responsible for line-shaped pits on (001) are nearly parallel to the (011) plane and therefore do not intersect the surface [9]. Furthermore, (011) epitaxial layers exhibit significantly reduced chlorine incorporation compared to (001), enabling low background carrier concentrations on the order of $10^{15}$ $cm^{-3}$, which is essential for high-breakdown-voltage devices [6, 10, 11].

Several studies have demonstrated the advantages of (011)-oriented β-$Ga_2O_3$. Hossain Sarkar et al. demonstrated high-quality (011) epitaxial layers grown by metalorganic chemical vapor deposition (MOCVD) with thicknesses up to 20 μm, exhibiting smooth surfaces, low threading dislocation densities, and low impurity concentration including H, C, and Si, even at a high growth rate of ~7.5 μm/h [12]. Oshima et al. reported rapid homoepitaxial growth on (011) substrates by HCl-based halide vapor phase epitaxy (HVPE), achieving growth rates up to ~14 μm/h while maintaining high crystalline quality and low impurity incorporation [11]. This approach effectively addresses the concern that epitaxial growth on (011) substrates is slower than that on (001), and represents a significant step toward the practical implementation of (011) epitaxy. As substrates for epitaxial growth, (011)-oriented 2-inch β-$Ga_2O_3$ wafers grown by the vertical Bridgman (VB) method have been demonstrated [13]. The development of the VB method addresses the limitation of edge-defined film-fed growth (EFG), which is restricted in the accessible crystal orientations. The (011) orientation also exhibits the smallest in-plane anisotropy in mechanical properties, such as the elastic modulus [14]. In addition, gas-phase etching processes tailored for (011) have been developed, enabling damage-free microfabrication with well-defined crystallographic features [15]. By utilizing a thick epilayer (~85 μm) with a low background carrier concentration ($1.8 \times 10^{15}$ $cm^{-3}$), a multi-fin vertical β-$Ga_2O_3$ transistor fabricated on a (011) substrate has been demonstrated, achieving a breakdown voltage exceeding 10 kV and a low specific on-resistance of 289 mΩ·$cm^2$ [6].

In parallel with the development of β-$Ga_2O_3$ substrates and epitaxial technologies, considerable efforts have been devoted to understanding the crystallographic nature of dislocations. Previous experimental and theoretical studies have shown that the dominant slip system in β-$Ga_2O_3$ is {001}⟨010⟩ [16, 17, 18, 19], although additional slip systems may become active depending on the stress state and crystal orientation [14, 20]. X-ray topography [4, 21] and transmission electron microscopy [22, 23, 24, 25, 26, 27] have revealed various extended defects, including threading dislocations, dislocation arrays associated with domain boundaries, and pipe-shaped voids. However, these studies have primarily focused on conventional substrate orientations such as (001), (010), and (100), which are typically fabricated from EFG-grown bulk crystals with the growth direction parallel to [010]. In contrast, the dislocation configurations and propagation behavior in (011)-oriented substrates grown in the direction normal to the (011) plane by the VB method remain largely unexplored.

Despite these advances in crystal growth and the current understanding of dislocation behavior in β-$Ga_2O_3$, several fundamental issues remain unresolved for (011)-oriented substrates. For example, although it has been suggested that dislocations responsible for line-shaped pits on (001) surfaces do not emerge on (011) due to crystallographic constraints [9], the detailed nature of these dislocations and the behavior of other dislocation types—such as those propagating along the *b*-axis (i.e., the [010] direction)—are not yet fully understood. In particular, it remains unclear whether the advantages of (011) epitaxy over (001), especially with respect to dislocation behavior, depend on the substrate growth method, such as VB versus EFG.

In this work, we address these issues by investigating dislocation behavior in (011)-oriented β-$Ga_2O_3$ substrates grown by the VB method using X-ray topography (XRT) in both transmission and reflection geometry, together with X-ray reticulography. By visualizing the three-dimensional propagation of dislocations, we reveal their characteristic alignment, which leads to the formation of dislocation arrays at domain boundaries. We further identify the crystallographic planes on which the dislocations lie and clarify their termination behavior at the (011) surface. In addition, we investigate dislocations parallel to the (011) plane, which, according to previous studies, are associated with line-shaped pits in (001) epitaxial layers. These results provide insight into defect formation mechanisms relevant to epitaxial growth and device performance.

The bulk crystal was grown by the VB method in an $N_2$+$O_2$ ambient using a Pt–Rh crucible. The growth direction was normal to the (011) plane, achieved using a (011)-oriented seed crystal. The crucible descent rate and rotation speed were 1 mm/h and 3 rpm, respectively. Further details of the growth conditions can be found elsewhere [**13**, **28**, **29**, **30**, **31**, **32**, **33**]. The substrates were sliced from the VB-grown bulk crystal. Mechanical polishing and CMP were performed on both sides of the substrates to enable clear dislocation observation during transmission XRT. The substrate thickness was approximately 460 μm.

XRT observations in transmission geometry, taking advantage of the dynamical diffraction phenomenon known as anomalous transmission (Borrmann effect) [**34**, **35**], were performed at beamline BL24XU of SPring-8 [**36**, **37**, **38**, **39**, **40**]. The diffraction vector $\boldsymbol{g}$ = $\bar{7}12$ was selected because it provides a strong two-beam diffraction condition under the Borrmann geometry, enabling high-contrast imaging of dislocations throughout the entire substrate thickness. Furthermore, according to Oshima [**41**], the ($\bar{7}12$) planes correspond to one of the low-index {110} face-centered cubic planes when the oxygen sublattice of β-$Ga_2O_3$ is regarded as a cubic close-packed lattice, making this reflection particularly suitable for the present study. The X-ray wavelength was λ = 0.124 nm. Reflection XRT measurements were carried out at beamline BL-3C of the Photon

Factory, High Energy Accelerator Research Organization (KEK) [42, 43], using reflections with $\boldsymbol{g}$ = $7\bar{3}\bar{2}$ (λ=0.107 nm, incident angle 5°) and $\boldsymbol{g}$ = $0\bar{2}\bar{2}$ (λ=0.134 nm, incident angle 30°). These reflections were selected because they provide suitable asymmetric and symmetric reflection geometries, respectively, together with sufficiently strong diffraction intensity and favorable diffraction conditions within the accessible X-ray wavelength range of the beamline, thereby enabling clear observation of the emergence points and contrast characteristics of near-surface dislocations. X-ray reticulography [44, 45, 46, 47], a highly sensitive technique for detecting domain boundaries, was performed based on reflection XRT with the insertion of a metallic mesh.

Figure 1(a) shows the (011) pole figure of β-$Ga_2O_3$, in which each spot represents a specific crystallographic orientation and can be interpreted as a Laue spot corresponding to a particular diffraction vector $\boldsymbol{g}$. For this orientation, the pattern exhibits relatively low symmetry due to the monoclinic crystal structure of β-$Ga_2O_3$. The Laue spots corresponding to the diffraction vector $\boldsymbol{g}$ = $\bar{7}12$, used for transmission XRT observations, are highlighted in red, together with the major crystallographic orientations 100, 010, and 001. The $[01\bar{1}]$ direction is oriented toward the right. It should be noted that although the (011) and $(0\bar{1}\bar{1})$ planes are crystallographically equivalent in monoclinic $C2/m$ space group, they are not equivalent in the context of the goniometer geometry used for XRT measurements. Without properly distinguishing between these two surfaces, the Bragg condition for the selected $\boldsymbol{g}$-vector cannot be satisfied. In the present configuration, the X-ray beam enters the crystal through the (011) surface and exits from the $(0\bar{1}\bar{1})$ surface.

Figure 1(b) shows a photograph of the sample and the fluorescent screen (FS) under the two-beam Borrmann-effect condition. The small bright spot indicated by the arrow in the lower right corresponds to the X-ray entrance point, which is visible due to X-ray-excited luminescence from point defects in the β-$Ga_2O_3$ crystal. The two larger bright spots represent the forward-diffracted (transmitted) and diffracted beams with $\boldsymbol{g} = \bar{7}12$ reaching the FS. After confirming that the intensities of the two beams were comparable (i.e., satisfying the two-beam approximation), the FS was moved upwards to allow the forward-diffracted beam to enter a two-dimensional detector positioned behind it (Figure 1(c)). This configuration enables acquisition of transmission XRT images containing dislocation contrast. Details of the detector are provided elsewhere [48].

Figure 2(a) shows the transmission XRT image over a 6.6 mm × 7.6 mm area, corresponding to a projected dislocation image viewed from the $(0\bar{1}\bar{1})$ surface. Isolated dislocations with various line directions and contrasts, as well as dislocation arrays, can be clearly observed. The surface termination points of the dislocation arrays are aligned along the [100] and [001] directions, denoted as DA type-I and type-II, respectively, in the figure. Details of DA type-I are shown in Figure 2(b). The dislocations in the array extend approximately along the [010] direction (i.e., the *b*-axis), and their projected lengths indicate that they penetrate the entire substrate. In Borrmann-effect XRT, where dynamical diffraction dominates the dislocation contrast, the portions of dislocations closer to the exit surface of the X-ray exhibit stronger contrast than those near the entrance surface [34, 38]. Therefore, the lower-right ends of the dislocations correspond to their emergence points on the $(0\bar{1}\bar{1})$ surface. The fringe-like features are characteristic signatures of dynamical X-ray diffraction, related to the extinction length of the X-ray [35]. The projected length of these dislocations provides important information for identifying the crystallographic plane on which they lie. Considering the transmission XRT geometry using the $\boldsymbol{g} = \bar{7}12$ reflection, the angle between the (001) plane and the $(0\bar{1}\bar{1})$ exit surface, together with the Bragg angle, predicts that dislocations extending along the [010] direction on the (001) plane should exhibit a projected length of approximately 55% of their actual length (Figure 1(d)) [49]. Assuming that these dislocations penetrate the entire substrate thickness (460 μm), the expected projected length is approximately 250 μm, in good agreement with the observed values. In contrast, other low-index planes belonging to the [100] zone axis cannot reproduce the observed projection geometry. Therefore, the observed dislocations are assigned to the (001) plane, forming dislocation walls. This interpretation is also consistent with the well-established {001}⟨010⟩ slip system in β-$Ga_2O_3$ [**16**, **17**, **18**, **19**, **43**]. Subsequent reticulography measurements further reveal that these arrays are associated with domain boundaries along the (001) plane. In contrast, DA type-II exhibits a considerably more complex morphology than DA type-I. To further investigate its structure, transmission XRT observations were performed from the same region using several diffraction vectors (Supplementary Figure S1). Burgers-vector analysis reveals that the DA type-II array exhibits a comb-like morphology, consisting of a main trunk and numerous inclined branches. The main trunk remains visible under $\boldsymbol{g}$ = $\bar{4}00$ and $\boldsymbol{g}$ = $00\bar{2}$, indicating dislocations possessing both $a$- and $c$-components in their Burgers vectors, where the $a$- and $c$-components correspond to the [100] and [001] directions, respectively. Such dislocations have previously been reported in EFG-grown β-$Ga_2O_3$ crystals [**38**, **42**]. In contrast, the inclined branches are invisible under these diffraction conditions but become visible under $\boldsymbol{g}$ = $0\bar{2}0$, indicating Burgers vectors dominated by the $b$-component ([010] direction). These observations demonstrate that DA type-II consists of two distinct dislocation populations with different Burgers vectors, suggesting a more complex formation mechanism than that of DA type-I. Most isolated dislocations outside the arrays exhibit line directions similar to those in DA type-I, i.e., lying on the (001) plane and extending along [010], although some show slight inclination or curvature. This suggests that the [010] direction remains a preferred propagation direction for dislocations during (011) VB growth, even though it is not perpendicular to the growth surface.

Figure 2(c) shows an area containing dislocations with long, thick contrast, typically extending along the [100] direction. In Borrmann-effect XRT, a dislocation exhibiting uniform contrast along its entire length indicates a line lying on a plane parallel to the sample surface. Therefore, these dislocations are attributed to lie on the (011) plane. Note that because transmission XRT provides a projection image through the entire substrate thickness, different dislocations may lie on different parallel (011) planes, and their depth positions cannot be distinguished from a single XRT image. Furthermore, they are invisible under $\boldsymbol{g} = \bar{4}01$ (Supplementary Figure S2), indicating that their Burgers vectors are most likely parallel to the [010] direction.

The presence of dislocations lying on the (011) plane is particularly interesting because Goto *et al*. [9] proposed that dislocations parallel to the (011) plane are responsible for the groove-like features formed on (001)-oriented homoepitaxial layers. They further reported that these groove-like features are greatly reduced when (011)-oriented substrates are used, suggesting that the corresponding dislocations are parallel to the (011) plane and therefore rarely intersect the (011) surface. Subsequent TEM observations confirmed that one such dislocation lies on the (011) plane and extends normal to the [100] direction [9]. These findings have led several studies to attribute the superior surface morphology and device performance of (011)-based epitaxial layers to the suppression of such groove-forming dislocations [50].

Since the present study also reveals the existence of dislocations lying on the (011) plane, it is important to determine whether they correspond to the same dislocation type reported by Goto *et al*. Our observations show that these dislocations predominantly extend along the [100] direction, although some locally deviate onto other crystallographic planes during crystal growth, as indicated by the red arrows in Figure 2(c). Their propagation direction therefore differs from that reported by Goto *et al*., indicating that they represent a different dislocation configuration. One possible origin of this difference is the different bulk crystal growth methods employed, namely EFG in Ref. [9] and the VB method in the present study, which may lead to different dislocation formation and propagation behaviors.

It should also be noted that the present study reveals another dominant dislocation population lying on the (001) plane and extending along the [010] direction. Unlike the dislocations discussed above, these dislocations intersect the (011) substrate surface and therefore may propagate into the epitaxial layer. Whether these surface-terminated dislocations generate surface features or influence the performance of (011)-oriented epitaxial devices remains an important subject for future investigation. Such information will be essential for evaluating the advantages and limitations of different substrate orientations for β-$Ga_2O_3$ power devices.

Reflection XRT provides information complementary to that obtained by transmission XRT. Whereas transmission XRT records the projection of the three-dimensional dislocation distribution through the entire substrate thickness, reflection XRT is sensitive only to the near-surface region and therefore identifies the dislocations terminating at the observed surface. Furthermore, because reflection XRT is performed under Bragg-case diffraction with a limited X-ray penetration depth, the dislocation contrast is dominated by orientation contrast and can be interpreted using kinematical diffraction theory. Consequently, the observed spot morphology can be directly compared with ray-tracing simulations to infer the dislocation character [51, 52]. In addition, the reflection XRT observations provide the basis for the subsequent reticulography measurements, allowing direct correlation between dislocation arrays and lattice misorientation across domain boundaries.

Figure 3 shows reflection XRT images observed from the $(0\bar{1}\bar{1})$ surface, acquired with $\boldsymbol{g}$ = $7\bar{3}\bar{2}$ from an area near that shown in Figure 2. The X-ray penetration depth is estimated to be several micrometers [53]. Bright spots corresponding to the emergence points of threading dislocations show good correlation with the transmission XRT images. Figures 3(b) and 3(c) present magnified views of two representative regions containing DA type-I arrays and isolated dislocations, respectively. It is evident that all spots within the arrays exhibit similar shape and contrast, indicating that these dislocations share the same character in terms of line direction and Burgers vector. In contrast, isolated spots show slightly higher intensity and larger size (Figure 3(b)), and all exhibit a head–tail feature, as indicated by the yellow circles in Figure 3(c). These observations suggest that the isolated dislocations differ from those in the arrays, and that their character can be evaluated and classified based on detailed analysis of spot contrast and size in reflection XRT. A study combining reflection XRT with ray-tracing simulations [42, 54] is currently in progress.

Figure 4(a) shows the reflection XRT image acquired using $\boldsymbol{g}$ = $0\bar{2}\bar{2}$, and Figure 4(b) shows the corresponding reticulographic image. Multiple arrays of bright spots, each extending along the [100] direction, are clearly observed. As transmission XRT has confirmed that DA type-I dislocations lie on the (001) plane, these arrays are most likely aligned along (001) domain boundaries. Such dislocations would not emerge on (001) substrates but do appear in the case of (011), potentially leading to a higher fraction of dislocations propagating from the substrate into the epilayer. In addition, the presence of domain boundaries may influence epitaxial growth because they are associated with local crystallographic discontinuities. Therefore, it is important to evaluate domain boundaries and take them into account when comparing (001) and (011) substrates. X-ray reticulography provides a powerful tool for this purpose [46]. As shown in Figures 4(b) and 4(c), the mesh aperture images exhibit characteristic deviations from an ideal periodic arrangement, indicating the presence of domain boundaries. Figure 4(e) schematically illustrates the distortion pattern observed in Figures 4(a) and 4(c). The red arrows indicate the relative vertical displacement of the diffraction spots across the domain boundary, which is characteristic of a detectable twist-type lattice misorientation. In addition, the enlarged view in Figure 4(d) shows that the spacing between adjacent spot columns changes from d to d', where d'>d. This distortion pattern is schematically illustrated in Figure 4(f) and is characteristic of a detectable tilt-type lattice misorientation. The magnitude of the detectable lattice misorientation was estimated by comparison with the calibrated reticulographic patterns reported previously [**46**], together with ray-tracing simulations using our in-house simulation code, yielding a typical value on the order of $10^{-5}$ rad. These observations indicate that the domain boundaries in the present VB-grown β-$Ga_2O_3$ substrate generally possess a mixed character, consisting of both detectable twist- and tilt-type lattice misorientation components, rather than being purely twist or purely tilt boundaries. This magnitude is in good agreement with the dislocation density in the arrays. For comparison, domain boundaries in EFG-grown (001) substrates are typically aligned along the (100) plane, with dislocation emergence points aligned along the [010] direction. This configuration may correspond to DA type-II observed in this study. The misorientation is likewise on the order of $10^{-5}$ rad [**46**].

In summary, dislocation behavior in (011)-oriented β-$Ga_2O_3$ substrates grown by the VB method was investigated using transmission and reflection XRT, complemented by X-ray reticulography. Transmission XRT revealed that most dislocations lie on the (001) plane and extend along [010], forming arrays associated with domain boundaries. Dislocations on the (011) plane extending along [100] were also identified and differ from those responsible for line-shaped pits on (001) epitaxial surfaces. Reflection XRT showed good agreement with transmission XRT and enabled classification of dislocation types based on contrast features. Reticulography confirmed domain boundaries with misorientation on the order of $10^{-5}$ rad, consistent with the dislocation density in the arrays. Furthermore, Burgers-vector analyses revealed that the dominant dislocations are characterized by Burgers vectors parallel to the [010] direction, whereas the more complex DA type-II arrays consist of dislocations possessing different Burgers-vector components.

These results indicate that dislocation behavior in (011) substrates is more complex than previously assumed and may depend on the substrate growth method, highlighting the importance of considering both crystallographic orientation and growth conditions when optimizing substrates for epitaxial growth.

**Supplementary Materials**

See Supplementary Figure S1 for transmission XRT images of a representative DA type-II array recorded using multiple diffraction vectors, which were used for the Burgers-vector analysis of the main trunk and inclined branches.

See Supplementary Figure S2 for the transmission XRT image recorded using the $\boldsymbol{g} = \bar{4}01$ reflection, demonstrating that most dislocations lying on the (011) plane are invisible under this diffraction condition.

## Acknowledgments

This study was supported by Innovative Science and Technology Initiative for Security Grant No. JPJ004596, ATLA, Japan, and JSPS KAKENHI Grant No. 20K05355, 23H01872, and 23K17356 Japan. The authors thank Novel Crystal Technology for preparing the samples. The synchrotron XRT observations were performed at BL-3C and BL-14B of KEK-PF under proposal Nos. 2020G585, 2022G503, and 2024G520, and at BL24XU of SPring-8 with approval from the Japan Radiation Research Institute (Proposal No. 2023A3055, 2023B3055, 2024A3055, 2024B3055). Y.Y. gratefully acknowledges Prof. Dr. K. Hirano and Prof. Dr. Y. Tsusaka for their support in the XRT experiments.

## AUTHOR DECLARATIONS

### Declaration of competing interests

The authors declare that they have no known competing financial interests or personal relationships that could have appeared to influence the work reported in this paper.

### Generative AI

Not used in the manuscript preparation process.

### Author contributions

Yongzhao Yao: Conceptualization (lead); Data curation (equal); Formal analysis (lead); Writing - original draft (lead); Writing - review & editing (equal).

Daiki Katsube: Data curation (equal); Formal analysis (equal); Writing - review & editing (equal).

Hirotaka Yamaguchi: Data curation (equal); Formal analysis (equal); Writing - review & editing (equal).

Yukari Ishikawa: Funding acquisition (lead); Project administration (lead); Writing - review & editing (equal).

All authors have approved the manuscript.

## Data Availability

Raw data were generated at the synchrotron facilities KEK-PF and SPring-8. The data that support the findings of this study are available within the article.

## REFERENCES

[1] S. Pearton, J. Yang, P. Cary, F. Ren, J. Kim, M. Tadjer, and M. Mastro, A review of β-$Ga_2O_3$ materials, processing, and devices, Appl. Phys. Rev. 5, 011301 (2018). [DOI:10.1063/1.5006941]

[2] M. Higashiwaki, β-$Ga_2O_3$ material properties, growth technologies, and devices: a review, AAPPS Bull. 32, 3 (2022). [DOI:10.1007/s43673-021-00033-0]

[3] S. Pearton, F. Ren, M. Tadjer, and J. Kim, Perspective: β-$Ga_2O_3$ for ultra-high power rectifiers and MOSFETS, J. Appl. Phys. 124, 220901 (2018). [DOI:10.1063/1.5062841]

[4] K. Sasaki, Prospects for β-$Ga_2O_3$: now and into the future, Appl. Phys. Express 17, 090101-1 (2024). [DOI:10.35848/1882-0786/ad6b73]

[5] W. Li, K. Nomoto, Z. Hu, T. Nakamura, D. Jena and H. G. Xing, "Single and multi-fin normally-off β-$Ga_2O_3$ vertical transistors with a breakdown voltage over 2.6 kV," 2019 IEEE International Electron Devices Meeting (IEDM), San Francisco, CA, USA, 2019, pp. 12.4.1-12.4.4. [Doi:10.1109/IEDM19573.2019.8993526]

[6] D. Wakimoto, C. Lin, K. Ema, Y. Ueda, H. Miyamoto, K. Sasaki, and A. Kuramata, A multi-fin normally-off β-$Ga_2O_3$ vertical transistor with a breakdown voltage exceeding 10 kV, Appl. Phys. Express 18, 106502 (2025). [DOI:10.35848/1882-0786/ae0d2a]

[7] A. Fiedler, R. Schewski, M. Baldini, Z. Galazka, G. Wagner, M. Albrecht, and K. Irmscher, Influence of incoherent twin boundaries on the electrical properties of β-$Ga_2O_3$ layers homoepitaxially grown by metal-organic vapor phase epitaxy, J. Appl. Phys. 122, 165701 (2017). [DOI:10.1063/1.4993748]

[8] T. Ngo, D. Le, J. Lee, S. Hong, J. Ha, W. Lee, and Y. Moon, Investigation of defect structure in homoepitaxial (-201) β-$Ga_2O_3$ layers prepared by plasma-assisted molecular beam epitaxy, J. Alloys Compd. 834, 155027 (2020). [DOI:10.1016/j.jallcom.2020.155027]

[9] K. Goto, H. Murakami, A. Kuramata, S. Yamakoshi, M. Higashiwaki, and Y. Kumagai, Effect of substrate orientation on homoepitaxial growth of β-$Ga_2O_3$ by halide vapor phase epitaxy, Appl. Phys. Lett. 120, 102102 (2022). [DOI:10.1063/5.0087609]

[10] K. Ema, C.-H. Lin, Y. Ueda, K. Sasaki, and A. Kuramata, Homo-epitaxial growth on the β-$Ga_2O_3$ (011) substrate by HVPE, Extended Abstracts, No. 20p-A22-4, the 85th Autumn Meeting, the Japan Society of Applied Physics, Niigata, Japan, Sept. 20th, 2024.

[11] Y. Oshima and T. Oshima, Rapid homoepitaxial growth of (011) β-$Ga_2O_3$ by HCl-based halide vapor phase epitaxy, Sci. Technol. Adv. Mater. 26, 2585551 (2025). [DOI:10.1080/14686996.2025.2585551]

[12] M. Hossain Sarkar, D. Yu, S. Alam, M. Hassan, J. Hwang, and H. Zhao, Metalorganic chemical vapor deposition of (011) β-$Ga_2O_3$ films with 20 μm drift layer, ACS Appl. Electron. Mater. 8, 1380 (2026). [DOI:10.1021/acsaelm.5c02627]

[13] Y. Ueda, T. Igarashi, K. Koshi, R. Sakaguchi, T. Chujo, R. Shinagawa, K. Sasaki, and A. Kuramata, Growth of 2-inch n-type β-$Ga_2O_3$ (011) single-crystal by VB method, Extended Abstracts, No. 7a-N201-3, the 86th Autumn Meeting, the Japan Society of Applied Physics, Nagoya, Japan, Sept. 7th, 2025.

[14] Y. Yao, Y. Sugawara, K. Sasaki, A. Kuramata, and Y. Ishikawa, Anisotropic mechanical properties of β-$Ga_2O_3$ single-crystal measured via angle-dependent nanoindentation using a Berkovich indenter, J. Appl. Phys. 134, 215106 (2023). [DOI:10.1063/5.0180389]

[15] T. Oshima and Y. Oshima, Near-vertical plasma-free HCl gas etching on (011) β-$Ga_2O_3$, Jpn. J. Appl. Phys. 64, 018003 (2025). [DOI:10.35848/1347-4065/ada706]

[16] Y. Yao, K. Mizuno, K. Ohnishi, Y. Ishikawa, M. Kitahara, T. Tomida, R. Murakami, V. Kochurikhin, L. Gushchina, K. Kamada, K. Kakimoto, and A. Yoshikawa, Synchrotron-radiation x-ray topography and reticulography of bulk β-$Ga_2O_3$ crystals grown by the cold crucible method, J. Appl. Phys. 139, 205101 (2026). [DOI:10.1063/5.0335858]

[17] Y. Yao, D. Katsube, H. Yamaguchi, S. Yamaguchi, D. Wakimoto, H. Miyamoto, and Y. Ishikawa, Three-dimensional visualization of lattice defects in β-$Ga_2O_3$ via synchrotron-radiation Borrmann-effect X-ray topo-tomography, Appl. Phys. Lett. 128, 231904 (2026). [DOI:10.1063/5.0339598]

[18] H. Yamaguchi, A. Kuramata, and T. Masui, Slip system analysis and x-ray topographic study on β-$Ga_2O_3$, Superlattice Microst. 99, 99 (2016). [DOI:10.1016/j.spmi.2016.04.030]

[19] Y. Yao, Y. Ishikawa, and Y. Sugawara, Slip planes in monoclinic β-$Ga_2O_3$ revealed from its {010} face via synchrotron X-ray diffraction and X-ray topography, Jpn. J. Appl. Phys. 59, 125501 (2020). [DOI:10.35848/1347-4065/abc1aa]

[20] Y. Wu, S. Gao, and H. Huang, The deformation pattern of single crystal β-$Ga_2O_3$ under nanoindentation, Mater. Sci. Semicond. Process. 71, 321 (2017). [DOI:10.1016/j.mssp.2017.08.019]

[21] S. Masuya, K. Sasaki, A. Kuramata, S. Yamakoshi, O. Ueda, and M. Kasu, Characterization of crystalline defects in β-$Ga_2O_3$ single crystals grown by edge-defined film-fed growth and halide vapor-phase epitaxy using synchrotron X-ray topography, Jpn. J. Appl. Phys. 58, 055501 (2019). [DOI:10.7567/1347-4065/ab0dba]

[22] O. Ueda, N. Ikenaga, K. Koshi, K. Iizuka, A. Kuramata, K. Hanada, T. Moribayashi, S. Yamakoshi, and M. Kasu, Structural evaluation of defects in β-$Ga_2O_3$ single crystals grown by edge-defined film-fed growth process, Jpn. J. Appl. Phys. 55, 1202BD (2016). [DOI:10.7567/JJAP.55.1202BD]

[23] O. Ueda, M. Kasu, and H. Yamaguchi, Structural characterization of defects in EFG- and HVPE-grown β-$Ga_2O_3$ crystals, Jpn. J. Appl. Phys. 61, 050101 (2022). [DOI:10.35848/1347-4065/ac4b6b]

[24] K. Ogawa, N. Ogawa, R. Kosaka, T. Isshiki, Y. Yao, and Y. Ishikawa, Three-dimensional observation of internal defects in a β-$Ga_2O_3$ (001) wafer using the FIB–SEM serial sectioning method, J. Electron. Mater. 49, 5190 (2020). [DOI:10.1007/s11664-020-08313-5]

[25] K. Ogawa, N. Ogawa, R. Kosaka, T. Isshiki, T. Aiso, M. Iyoki, Y. Yao, and Y. Ishikawa, AFM observation of etch-pit shapes on β-$Ga_2O_3$ (001) surface formed by molten alkali etching, Mater. Sci. Forum 1004, 512 (2020). [DOI:10.4028/www.scientific.net/MSF.1004.512]

[26] K. Nakai, T. Nagai, K. Noami, and T. Futagi, Characterization of defects in β-$Ga_2O_3$ single crystals, Jpn. J. Appl. Phys. 54, 051103 (2015). [DOI:10.7567/JJAP.54.051103]

[27] S. Isaji, I. Maeda, N. Ogawa, R. Kosaka, N. Hasuike, T. Isshiki, K. Kobayashi, Y. Yao, and Y. Ishikawa, Relationship between propagation angle of dislocations in β-$Ga_2O_3$ (001) bulk wafers and their etch pit shapes, J. Electron. Mater. 52, 5093 (2023). [DOI:10.1007/s11664-023-10363-4]

[28] E. Ohba, T. Kobayashi, M. Kado, and K. Hoshikawa, Defect characterization of β-$Ga_2O_3$ single crystals grown by vertical Bridgman method, Jpn. J. Appl. Phys. 55, 1202BF (2016). [DOI:10.7567/JJAP.55.1202BF]

[29] K. Hoshikawa, E. Ohba, T. Kobayashi, J. Yanagisawa, C. Miyagawa, and Y. Nakamura, Growth of β-$Ga_2O_3$ single crystals using vertical Bridgman method in ambient air, J. Cryst. Growth 447, 36 (2016). [DOI:10.1016/j.jcrysgro.2016.04.022]

[30] K. Hoshikawa, T. Kobayashi, Y. Matsuki, E. Ohba, and T. Kobayashi, 2-inch diameter (1 0 0) β-$Ga_2O_3$ crystal growth by the vertical Bridgman technique in a resistance heating furnace in ambient air, J. Cryst. Growth 545, 125724 (2020). [DOI:10.1016/j.jcrysgro.2020.125724]

[31] K. Hoshikawa, T. Kobayashi, E. Ohba, and T. Kobayashi, 50 mm diameter Sn-doped (001) β-$Ga_2O_3$ crystal growth using the vertical Bridgeman technique in ambient air, J. Cryst. Growth 546, 125778 (2020). [DOI:10.1016/j.jcrysgro.2020.125778]

[32] E. Ohba, T. Kobayashi, T. Taishi, and K. Hoshikawa, Growth of (100), (010) and (001) β-$Ga_2O_3$ single crystals by vertical Bridgman method, J. Cryst. Growth 556, 125990 (2021). [DOI:10.1016/j.jcrysgro.2020.125990]

[33] T. Igarashi, Y. Ueda, K. Koshi, R. Sakaguchi, S. Watanabe, S. Yamakoshi, and A. Kuramata, Growth of 6 inch diameter β-$Ga_2O_3$ crystal by the vertical Bridgman method, Phys. Status Solidi B 262, 2400444 (2025). [DOI:10.1002/pssb.202400444]

[34] G. Borrmann, Die Absorption von Röntgenstrahlen im Fall der Interferenz, Z. Phys. 127, 297 (1950). [DOI:10.1007/BF01329828]

[35] A. Authier, Dynamical theory of X-ray diffraction, revised edition, Oxford University Press Inc., New York, (2001) p.139–p.147

[36] Y. Tsusaka, H. Mizuochi, M. Imanishi, M. Imade, Y. Mori, and J. Matsui, Identification of dislocation characteristics in Na-flux-grown GaN substrates using bright-field X-ray topography under multiple-diffraction conditions, J. Appl. Phys. 125, 125105 (2019). [DOI:10.1063/1.5082990]

[37] Y. Yao, K. Hirano, Y. Sugawara, K. Sasaki, A. Kuramata, and Y. Ishikawa, Observation of dislocations in thick β-$Ga_2O_3$ single-crystal substrates using Borrmann effect synchrotron x-ray topography, APL Mater. 10, 051101 (2022). [DOI:10.1063/5.0088701]

[38] Y. Yao, Y. Tsusaka, K. Sasaki, A. Kuramata, Y. Sugawara, and Y. Ishikawa, Large-area total-thickness imaging and Burgers vector analysis of dislocations in β-$Ga_2O_3$ using bright-field x-ray topography based on anomalous transmission, Appl. Phys. Lett. 121, 012105 (2022). [DOI:10.1063/5.0098942]

[39] Y. Yao, Y. Tsusaka, K. Hirano, K. Sasaki, A. Kuramata, Y. Sugawara, and Y. Ishikawa, Three-dimensional distribution and propagation of dislocations in β-$Ga_2O_3$ revealed by Borrmann effect X-ray topography, J. Appl. Phys. 134, 155104 (2023). [DOI:10.1063/5.0169526]

[40] Y. Yao, Y. Tsusaka, and Y. Ishikawa, Kinematical and dynamical contrast of dislocations in thick GaN substrates observed by synchrotron-radiation x-ray topography under six-beam diffraction conditions, J. Appl. Phys. 139, 065102 (2026). [DOI:10.1063/5.0314775]

[41] T. Oshima, Mapping primary crystallographic planes in β-$Ga_2O_3$ based on a pseudo-cubic oxygen sublattice, Jpn. J. Appl. Phys. 65, 038003 (2026). [DOI:10.35848/1347-4065/ae42ac]

[42] Y. Yao, Y. Sugawara, and Y. Ishikawa, Identification of Burgers vectors of dislocations in monoclinic β-$Ga_2O_3$ via synchrotron x-ray topography, J. Appl. Phys. 127, 205110 (2020). [DOI:10.1063/5.0007229]

[43] Y. Yao, D. Wakimoto, H. Miyamoto, K. Sasaki, A. Kuramata, K. Hirano, Y. Sugawara, and Y. Ishikawa, X-ray topographic observation of dislocations in β-$Ga_2O_3$ Schottky barrier diodes and their glide and multiplication under reverse bias, Scr. Mater. 226, 115216 (2023). [DOI:10.1016/j.scriptamat.2022.115216]

[44] A. Lang and A. Makepeace, Synchrotron x-ray reticulography: principles and applications, J. Phys. D: Appl. Phys. 32, A97 (1999). [DOI:10.1088/0022-3727/32/10A/321]

[45] R. Arridge, A. Lang, and A. Makepeace, Elastic deformation in a crystal plate where lattice–parameter mismatch is present between adjacent growth sectors. II. Measurement of lattice tilts by synchrotron X–ray reticulography, Proc. R. Soc. Lond. A 458, 2623 (2002). [DOI:10.1098/rspa.2002.0998]

[46] Y. Yao, K. Hirano, K. Sasaki, A. Kuramata, Y. Sugawara, and Y. Ishikawa, Lattice misorientation at domain boundaries in β-$Ga_2O_3$ single-crystal substrates observed via synchrotron radiation X-ray diffraction imaging and X-ray reticulography, J. Am. Ceram. Soc. 106, 5487 (2023). [DOI:10.1111/jace.19156]

[47] Y. Yao, K. Hirano, Y. Sugawara, and Y. Ishikawa, Domain boundaries in ScAlMgO4 single crystal observed by synchrotron radiation X-ray topography and reticulography, Semicond. Sci. Technol. 37, 115009 (2022). [DOI:10.1088/1361-6641/ac974b]

[48] Y. Yao, Y. Sugawara, Y. Ishikawa, and K. Hirano, X-ray topography of crystallographic defects in wide-bandgap semiconductors using a high-resolution digital camera, Jpn. J. Appl. Phys. 60, 010908 (2021). [DOI:10.35848/1347-4065/abd2dd]

[49] K. Momma and F. Izumi, VESTA 3 for three-dimensional visualization of crystal, volumetric and morphology data, J. Appl. Crystallogr. 44, 1272 (2011). [DOI:10.1107/S0021889811038970]

[50] S. Sdoeung, Y. Otsubo, K. Sasaki, A. Kuramata, and M. Kasu, Killer defect responsible for reverse leakage current in halide vapor phase epitaxial (011) β-$Ga_2O_3$ Schottky barrier diodes investigated via ultrahigh sensitive emission microscopy and synchrotron x-ray topography, Appl. Phys. Lett. 123, 122101 (2023). [DOI:10.1063/5.0170398]

[51] M. Dudley, X.R. Huang, and W. Huang, Assessment of orientation and extinction contrast contributions to the direct dislocation image, J. Phys. D: Appl. Phys. 32, A139 (1999). [DOI:10.1088/0022-3727/32/10a/329]

[52] B. Raghothamachar, G. Dhanaraj, J. Bai, and M. Dudley, Defect analysis in crystals using X-ray topography, Microsc. Res. Tech. 69, 343 (2006). [DOI:10.1002/jemt.20290]

[53] K. Ishiji, S. Kawado, Y. Hirai, and S. Nagamachi, Determination of observable depth of dislocations in 4H-SiC by X-ray topography in back reflection, Jpn. J. Appl. Phys. 56, 106601 (2017). [DOI:10.7567/JJAP.56.106601]

[54] M. Dudley, X. Huang, and W. Vetter, Contribution of x-ray topography and high-resolution diffraction to the study of defects in SiC, J. Phys. D: Appl. Phys. 36, A30 (2003). [DOI:10.1088/0022-3727/36/10a/307]

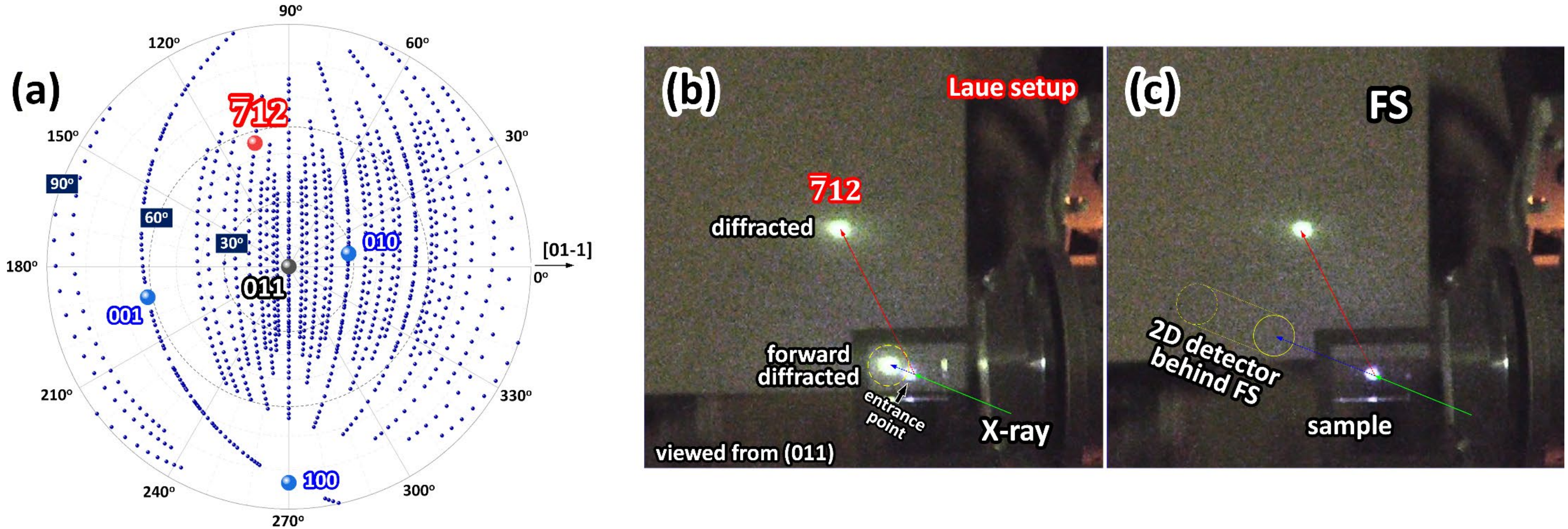

**Figure 1** (a) (011) pole figure of β-$Ga_2O_3$. (b),(c) Photograph of the sample and fluorescent screen (FS) under the two-beam Borrmann condition.

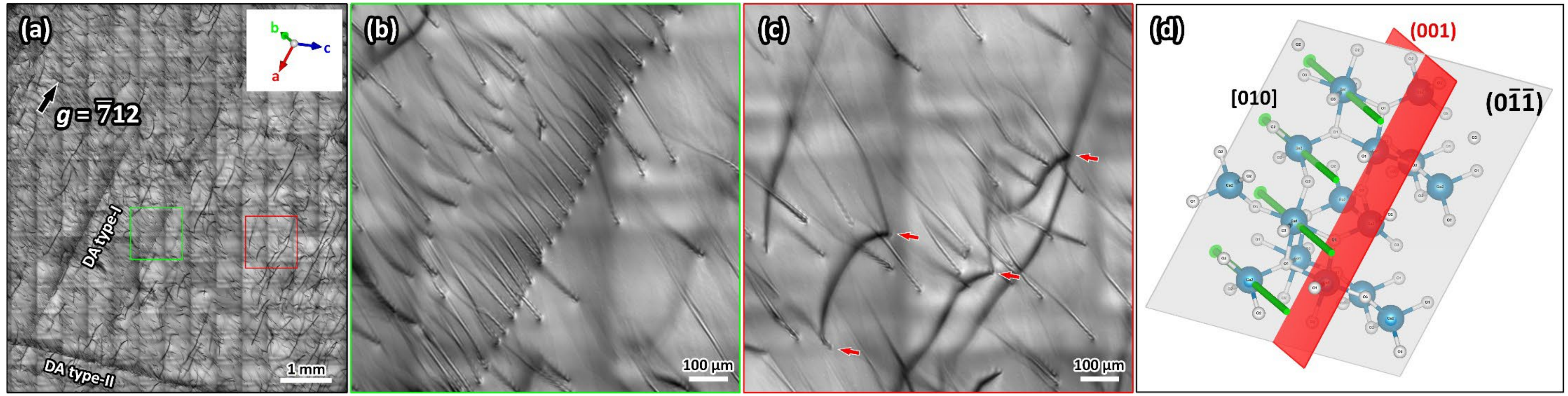

**Figure 2** (a) Transmission XRT image showing isolated dislocations and dislocation arrays. (b) DA type-I array aligned along [100], with dislocations extending along [010] on the (001) plane. (c) Dislocations with uniform thick contrast extending along [100], indicating lines parallel to the surface. (d) Schematic illustration of the crystal geometry. The gray and red planes represent the (0$\bar{1}\bar{1}$) and (001) planes, respectively. The green arrows indicate the [010] direction.

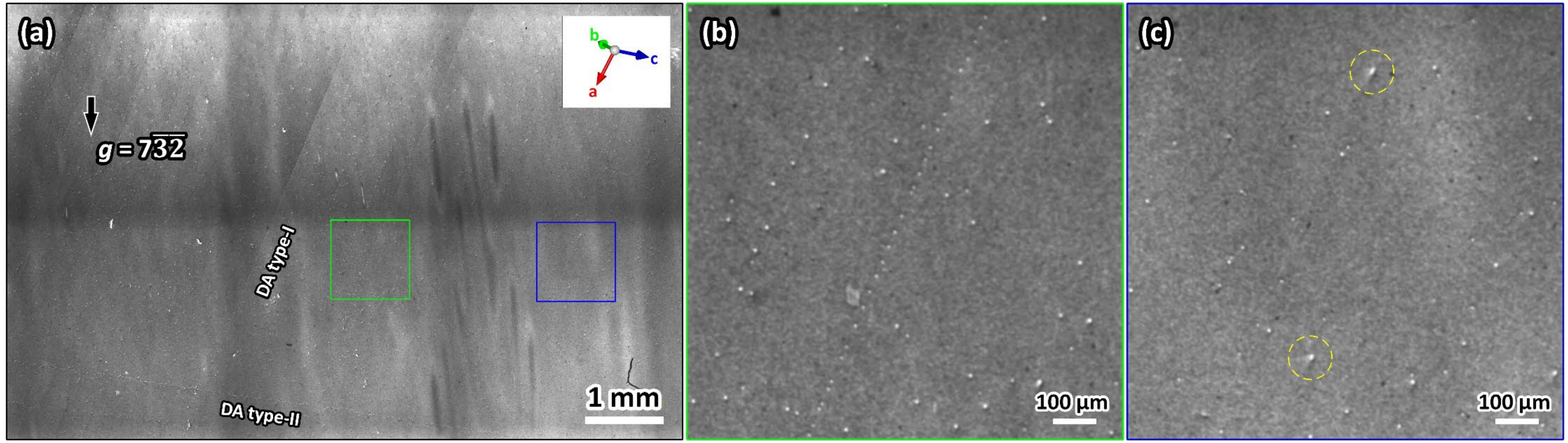

**Figure 3** (a) Reflection XRT image. (b) DA type-I array with uniform contrast. (c) Isolated dislocations showing larger size and head–tail contrast features.

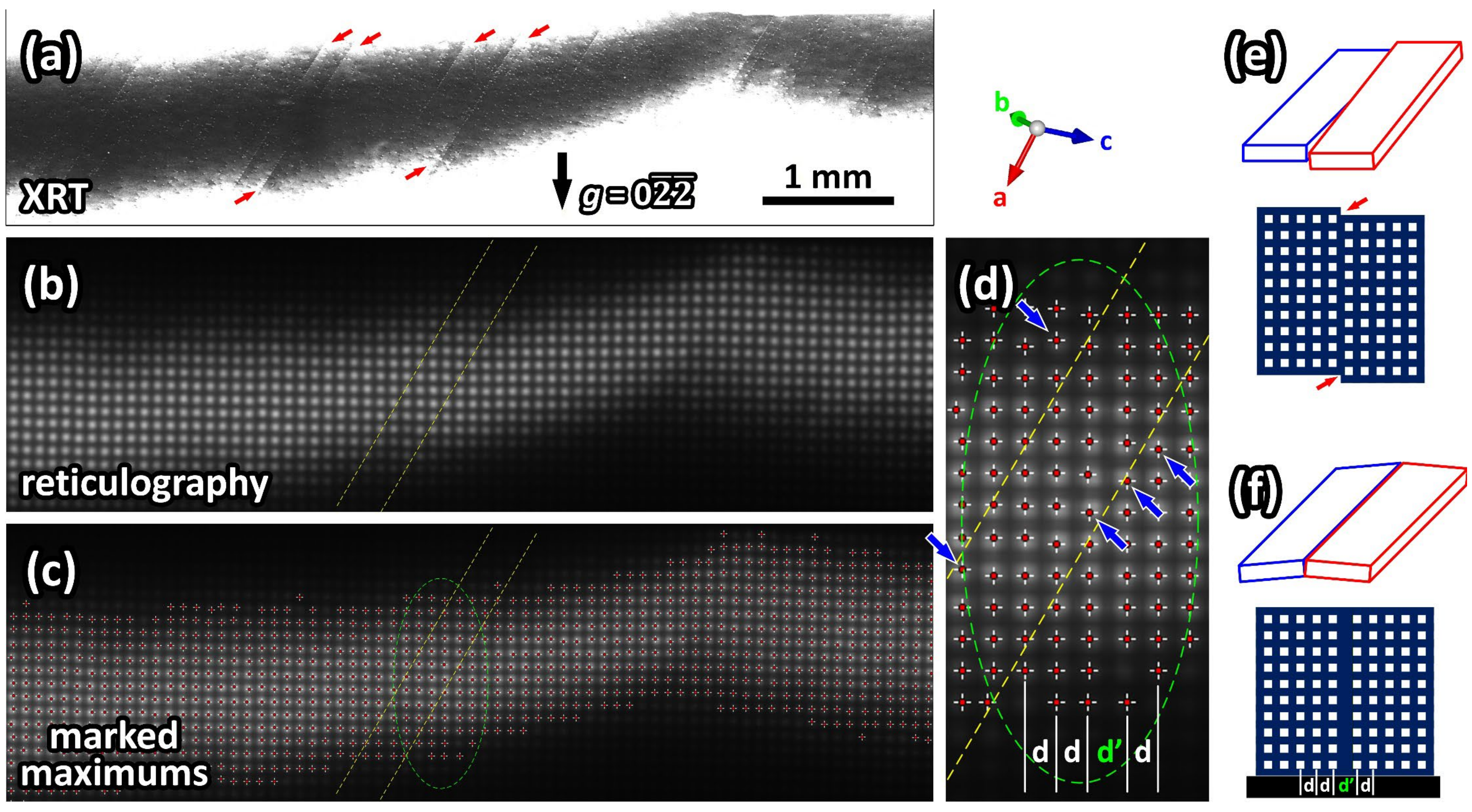

**Figure 4** (a) Reflection XRT image showing dislocation arrays along [100]. (b) Corresponding reticulography image. (c),(d) Maximum-intensity images of mesh apertures showing distortions indicative of domain boundaries with misorientation on the order of $10^{-5}$ rad. (e) Schematic illustration of the detectable twist-type lattice misorientation. (f) Schematic illustration of the detectable tilt-type lattice misorientation.

# Supplementary Materials

## Dislocations in (011)-oriented vertical Bridgman β-$Ga_2O_3$ substrates

Yongzhao Yao,[1,2,a)] Daiki Katsube,[2] Hirotaka Yamaguchi,[2] Yukari Ishikawa[2]

[1]Mie University, 1577 Kurimamachiya-cho, Tsu, Mie 514-8507, Japan
[2]Japan Fine Ceramics Center, 2-4-1 Mutsuno, Atsuta, Nagoya 456-8587, Japan

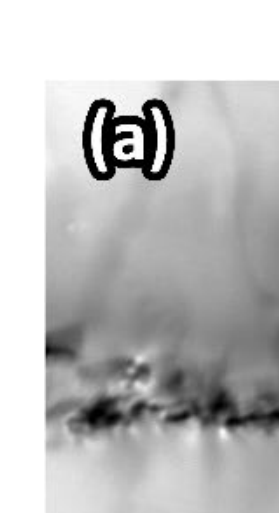
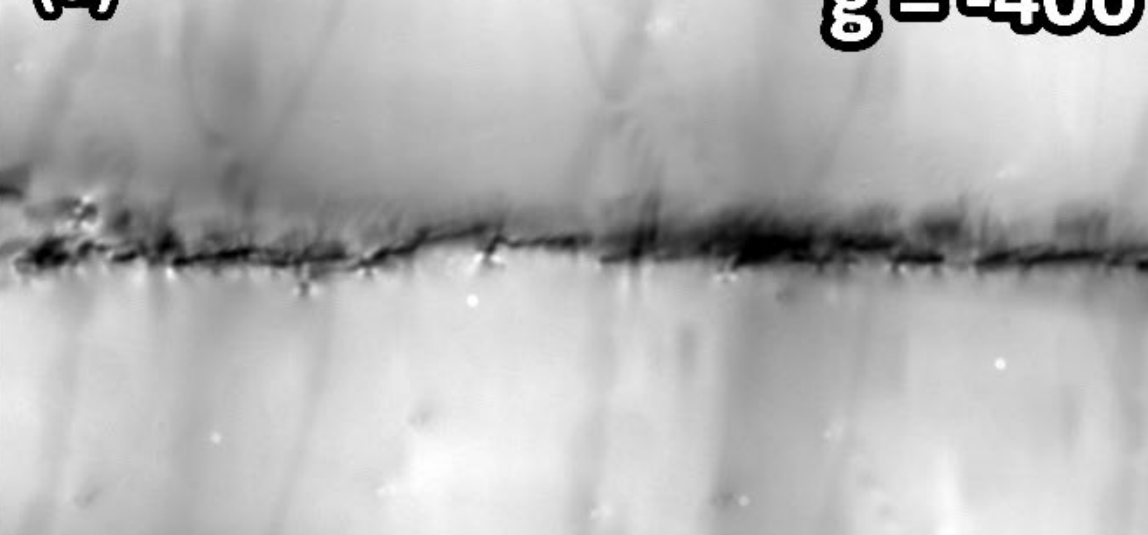

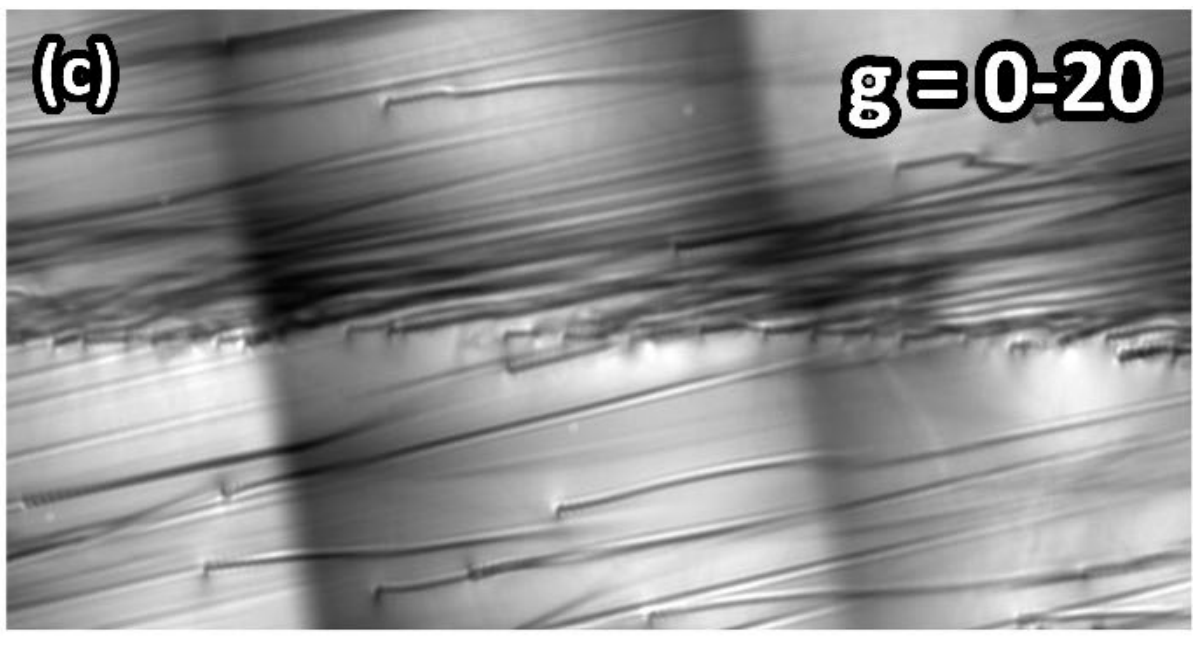

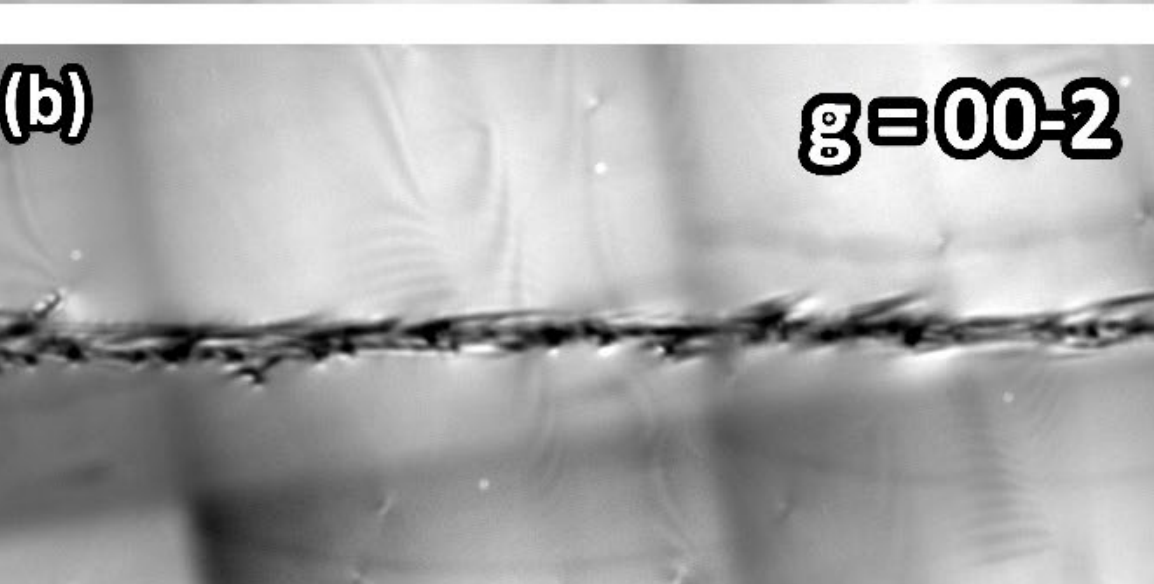

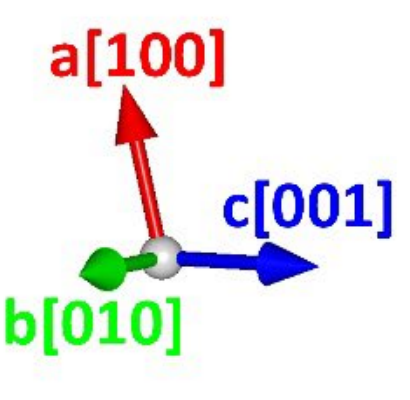

Figure S1 Transmission X-ray topography (XRT) images of the same DA type-II array acquired using different diffraction vectors for Burgers-vector analysis: (a) $\boldsymbol{g} = \bar{4}00$, (b) $\boldsymbol{g} = 00\bar{2}$, (c) $\boldsymbol{g} = 0\bar{2}0$. The DA type-II array exhibits a comb-like morphology, consisting of a main trunk and numerous inclined branches. The main trunk remains visible under $\boldsymbol{g} = \bar{4}00$ and $\boldsymbol{g} = 00\bar{2}$, indicating dislocations possessing both $a$- and $c$-components in their Burgers vectors, whereas the inclined branches are invisible under these diffraction conditions but become visible under $\boldsymbol{g} = 0\bar{2}0$, indicating Burgers vectors dominated by the $b$-component. Here, the $a$-, $b$-, and $c$-components correspond to the [100], [010], and [001] crystallographic directions, respectively.

ALT Text, Supplementary Figure S1:
Transmission X-ray topography images of a representative DA type-II dislocation array recorded under multiple diffraction conditions. The image set is used to identify the Burgers-vector characteristics of the main trunk and inclined branches.

[a)]Author to whom correspondence should be addressed. Electronic mail: yao@icsdf.mie-u.ac.jp.
ORCID:0000-0002-7746-4204

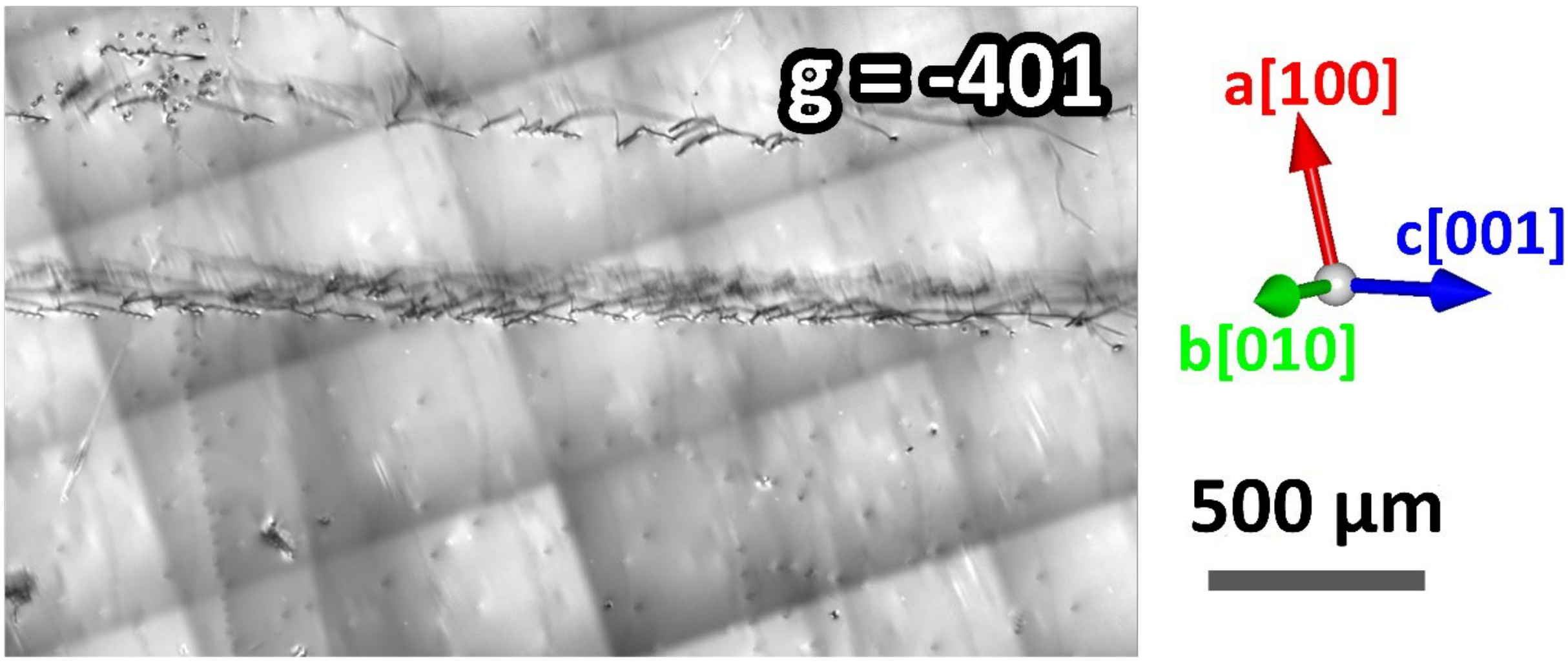

Figure S2 Transmission XRT image recorded using the $\boldsymbol{g} = \bar{4}01$ reflection. The dislocations lying on the (011) plane shown in Figure 2(c) are invisible under this diffraction condition, indicating that their Burgers vectors are most likely parallel to the [010] direction.

ALT Text, Supplementary Figure S2:

Transmission X-ray topography image recorded using the g=-401 reflection. Most dislocations lying on (011) planes are invisible under this diffraction condition, providing information on their Burgers-vector direction through the diffraction invisibility criterion.